\documentclass[aps,prl,floatfix,superscriptaddress,twocolumn,preprintnumbers]{revtex4} 
\usepackage{amssymb}
\usepackage{amsmath}
\usepackage{amsfonts}
\usepackage{epsfig}             
\usepackage{graphicx}
\usepackage{stackrel}
\usepackage{tabularx}
\usepackage{orcidlink}
\usepackage{color}
\usepackage{color}
\usepackage{dsfont}
\usepackage[T1]{fontenc}

\definecolor{darkerblue}{rgb}{0.0,0.0,0.5}
\newcommand{\seq}{\begin{subequations}}
\newcommand{\sen}{\end{subequations}}

\newcommand{\eq}{\begin{eqnarray}}
\newcommand{\en}{\end{eqnarray}}

\newcommand{\otm}{oT\!M}
\newcommand{\ee}{e^+ e^-}

\def\nn{\nonumber}

\begin{document}

\title{Formation of true muonium in the Drell-Yan dimuon production}

\author{Sergei~N.~Gninenko\,\orcidlink{0000-0001-6495-7619}}
\affiliation{Institute for Nuclear Research of the Russian Academy 
  of Sciences, 117312 Moscow, Russia}
  \affiliation{Millennium Institute for Subatomic Physics at  
the High-Energy Frontier (SAPHIR) of ANID, \\ 
	Fern\'andez Concha 700, Santiago, Chile}

\author{Sergey~Kuleshov\,\orcidlink{0000-0002-3065-326X}}
\affiliation{Millennium Institute for Subatomic Physics at  
the High-Energy Frontier (SAPHIR) of ANID, \\ 
	Fern\'andez Concha 700, Santiago, Chile}
\affiliation{Center for Theoretical and Experimental Particle Physics,
  Facultad de Ciencias Exactas, Universidad Andres Bello,
  Fernandez Concha 700, Santiago, Chile}

\author{Valery~E.~Lyubovitskij\,\orcidlink{0000-0001-7467-572X}}
\affiliation{Institut f\"ur Theoretische Physik,
Universit\"at T\"ubingen, \\
Kepler Center for Astro and Particle Physics, \\ 
Auf der Morgenstelle 14, D-72076 T\"ubingen, Germany}
\affiliation{Millennium Institute for Subatomic Physics at 
the High-Energy Frontier (SAPHIR) of ANID, \\
Fern\'andez Concha 700, Santiago, Chile}

\author{Alexey~S.~Zhevlakov\,\orcidlink{0000-0002-7775-5917}}
\affiliation{Bogoliubov Laboratory of Theoretical Physics, JINR, Dubna} 
\affiliation{Matrosov Institute for System Dynamics and 
 Control Theory SB RAS, \\  Lermontov str., 134,
 664033, Irkutsk, Russia } 
\date{\today}

\begin{abstract}

We suggest a new mechanism for the formation of  true muonium ($T\!M$),
the QED $(\mu^+ \mu^-)$ bound state,  due to the Coulomb interaction
of muons from  $\mu^+ \mu^-$ pairs generated in the
Drell-Yan reaction of protons  off nuclei of a thin foil target.
The  $T\!M$ atoms produced in the $^3S_1$ long-lived triplet state ($oT\!M$) 
could escape from the foil to vacuum followed by their decay
into $\ee$ pairs. Observation of  $\ee$ events with a   displaced vertex,
and  an  invariant mass and  decay time consistent with the mass and
lifetime of the $oT\!M$  is used as a signature of  the discovery of $T\!M$.   
We present estimates of the $T\!M$  yield and briefly discuss an 
experiment which could lead to such a discovery.

\end{abstract}

\maketitle

Leptonic atoms, which are QED bound states of two leptons with
opposite electric charge, such as positronium $(e^+ e^-)$,
muonium $(\mu^+ e^-)$, true muonium $(\mu^+ \mu^-)$,
true tauonium $(\tau^+ \tau^-)$, offer 
a unique possibility for the precision test
of QED~\cite{Karshenboim:2005iy,Brodsky:2009gx,Adkins:2022omi}.  
Since the Bohr radius $r_B$ of such atoms is quite large, e.g., $r_B \simeq 30$
fm for the most compact state - true tauonium,  and $r_B \simeq 512$ fm
for true muonium, they are bounded dominantly by the static Coulomb force.
Therefore, the properties of leptonic atoms (masses and decay rates)
are defined by their Coulomb wave functions at the origin.
On the other hand, QED can provide accurate inclusion of higher-order
corrections up to a desired order.

Complementary to leptonic atoms, we have strong
experimental and theoretical evidence of existence of
hadronic atoms~\cite{Bilenky:1969zd,Nemenov:1984cq,Gasser:2007zt},
which are bound states of oppositely charged hadrons, such, e.g., as 
pionium $(\pi^+ \pi^-)$, $(\pi^\pm K^\mp)$ atoms,  
pionic hydrogen $(\pi^- p)$, which also held together due
to static Coulomb interaction potential. Study of hadronic atoms serves
as powerful tool for independent check of predictions of QCD and QED. 
\par Study of leptonic atoms started with the theoretical analysis of
positronium in~\cite{Pirenne:1946pi} and its experimental 
observation~\cite{Deutsch:1951zza}.
A few years later the existence of the  second leptonic atom, muonium,
was theoretically predicted~\cite{Friedman:1957mz} and experimentally
confirmed~\cite{Hughes:1960zz}. Ideas for production and decay of
true muonium ($T\!M$) have been proposed in Ref.~\cite{Bilenky:1969zd}. 
In particular, it was suggested to produce $T\!M$ in the charge-exchange
reaction $\pi^- + p \to (\mu^+ \mu^-) + n$ and electroproduction off
nuclei $\gamma + Z \to  (\mu^+ \mu^-) + Z$. The accurate measure and
comparison of properties of $T\!M$ as its lifetime, branching fractions
of its decay modes, etc., with the SM predictions have a huge interest,
but up to now there is no even experimental evidence of $T\!M$ and
it is a large experimental challenge.

\par Many production mechanisms of
this leptonic atom, such as rare radiative meson decays 
$\eta(\eta',K_L) \to (\mu^+ \mu^-)
\gamma$~\cite{Nemenov:1972ph,Ji:2017lyh,Ji:2018dwx}, 
$e^+ e^-$~\cite{Moffat:1975uw,Brodsky:2009gx} 
and $\mu^+ \mu^-$~\cite{Hughes:1971mu} collisions, 
collisions of electrons with atoms 
$e + Z  \to e + (\mu^+ \mu^-) + Z$~\cite{Holvik:1986ty},
relativistic heavy ion collisions
$Z_1 + Z_2 \to Z_1 + Z_2 + (\mu^+ \mu^-)$~\cite{Ginzburg:1998df},
fixed-target (FT) experiments~\cite{Banburski:2012tk},
astrophysical sources (micro-quasar jet-star interactions, accretion
discs of both active galactic nuclei and micro-quasars)~\cite{Ellis:2015eea}
have been proposed. These results in several  planned  experiments on the $T\!M$ observation considered at
(1)~lepton colliders ($\mu\mu$-tron machine at BINP~\cite{Bogomyagkov:2017uul},
DIMUS at Fermilab~\cite{Fox:2021mdn}),  
(2)~hadron/heavy ion colliders (LHCb at CERN~\cite{CidVidal:2019qub}),
(3)~FT  experiments (REDTOP at Fermilab~\cite{Gatto:2016rae})
using neutral meson decays as a source of $T\!M$,
(4)~astrophysical sources~\cite{Ellis:2015eea}.  
In particular, the experiment at BINP~\cite{Bogomyagkov:2017uul} 
is based on the idea of Ref.~\cite{Brodsky:2009gx} 
to produce $T\!M$ in the $e^+e^-$ collisions with the energy near
the $T\!M$ mass. The DIMUS experiment at Fermilab~\cite{Gatto:2016rae}
will use the $e^+ e^-$ collider with the beam energy of 408 MeV. 
There are also plans to study the $T\!M$ production in the rare decays of
$\eta$ and $\eta'$ in the LHCb~\cite{CidVidal:2019qub} experiment. 

In this letter, we discuss a possibility for discovering of  $T\!M$  
in a FT experiment using a proton beam 
scattered off nuclei of a thin foil target.
Such collisions initiate Drell-Yan (DY) 
reaction with production of photon, which mixes with  the triplet $T\!M$, the so-called ortho- true muonium ($\otm$) - 
the $1^3S_1$ bound state, by analogy of the photon mixing with
the  Dark Photon ($A'$)~\cite{Holdom:1985ag}.
Note, that the such mixing has been previously
discussed in Ref.~\cite{CidVidal:2019qub}.

We consider $oT\!M$,  with the mass
$m_{_{oT\!M}} = 2m_\mu - E_b \simeq 211 $ MeV, where $m_\mu$
is the muon mass and $E_b = m_\mu \alpha^2/4 = 1.41 $ keV
is the $oT\!M$ binding energy, estimated for
the static Coloumb potential. 
We introduce $oT\!M$ as elementary field $M_\mu$ and
propose its mixing with QED photon $A_\mu$ following analogy
with Dark Photon~\cite{Holdom:1985ag,CidVidal:2019qub}: 
$\mathcal{L}_{\rm mix} = (\epsilon_{_{oT\!M}}/2) \, F_{\mu\nu} M^{\mu\nu}$, 
where $\epsilon_{_{oT\!M}}$ is the mixing parameter,
$F_{\mu\nu}$ and $M_{\mu\nu}$ are the stress tensors
of photon and $oT\!M$. 
Then making the shift of photon field
$A_\mu\to A_\mu - \epsilon_{_{oT\!M}} M_\mu$
we generate the coupling of $oT\!M$ with Standard Model (SM) fermions $\psi$: 
\eq\label{Lint}
\mathcal{L}_{int}=e \, \epsilon_{_{oT\!M}} 
M_\mu \sum_\psi \bar\psi
\gamma^\mu Q_\psi \psi \,, 
\en
where $Q_\psi$ is the charge of the SM fermion with flavor $\psi$. 

Mixing parameter $\epsilon_{_{oT\!M}}$ was fixed in Ref.~\cite{CidVidal:2019qub}
using QED result for the decay rate for the ground-state $oT\!M$
into $e^+ e^-$ pair, $\Gamma(oT\!M \to e^+ e^-) = \alpha^5 m_{_{oT\!M}}/12$,  
and the same quantity calculated using interaction Lagrangian~(\ref{Lint}):
$\Gamma(oT\!M \to e^+ e^-) = \alpha \, \epsilon_{_{oT\!M}}^2 \, m_{_{oT\!M}}/3$
(here we drop term of order $m_e^2/m_{_{oT\!M}}$).  
Matching two results for $\Gamma(oT\!M \to e^+ e^-)$ 
one gets $\epsilon_{_{oT\!M}}  = \alpha^2/2$~\cite{CidVidal:2019qub}. 
        
\begin{figure}[hb!]

 \vspace*{-1.55cm}  
 \hspace*{-1cm}
 	\includegraphics[scale=.7]{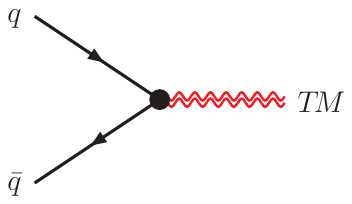}

 \vspace*{-17.6cm}                        
 \caption{LO partonic-level $q\bar q$ annihilation diagram
         contributing to the $T\!M$ production.}
\label{qqTM}

 \vspace*{-.7cm}                

 \hspace*{-3.2cm}
    	\includegraphics[scale=.7]{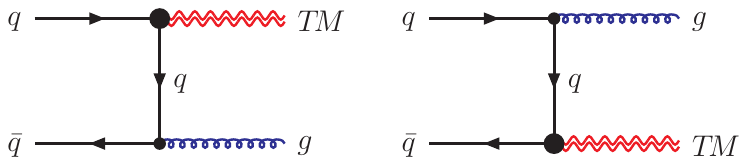} 

        \vspace*{-18.4cm}
	\caption{NLO partonic-level $q\bar q$ annihilation diagrams
		contributing to the $T\!M$ production.}
	\label{Diag_qq}
	
	\vspace*{-.6cm}
	
         \hspace*{-3.2cm}        
        	\includegraphics[scale=.7]{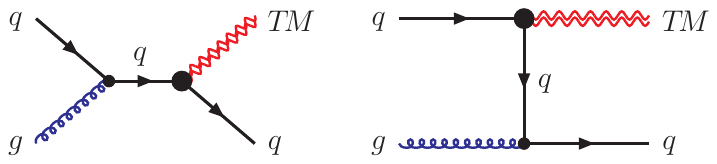}
                \vspace*{-18.7cm}	        
                \caption{NLO partonic-level $q g$ Compton scattering diagrams
                contributing to the $T\!M$ production.}
		\label{Diag_qg}	
\end{figure}

\begin{figure}[t]

  \includegraphics[width=0.48\textwidth,clip]{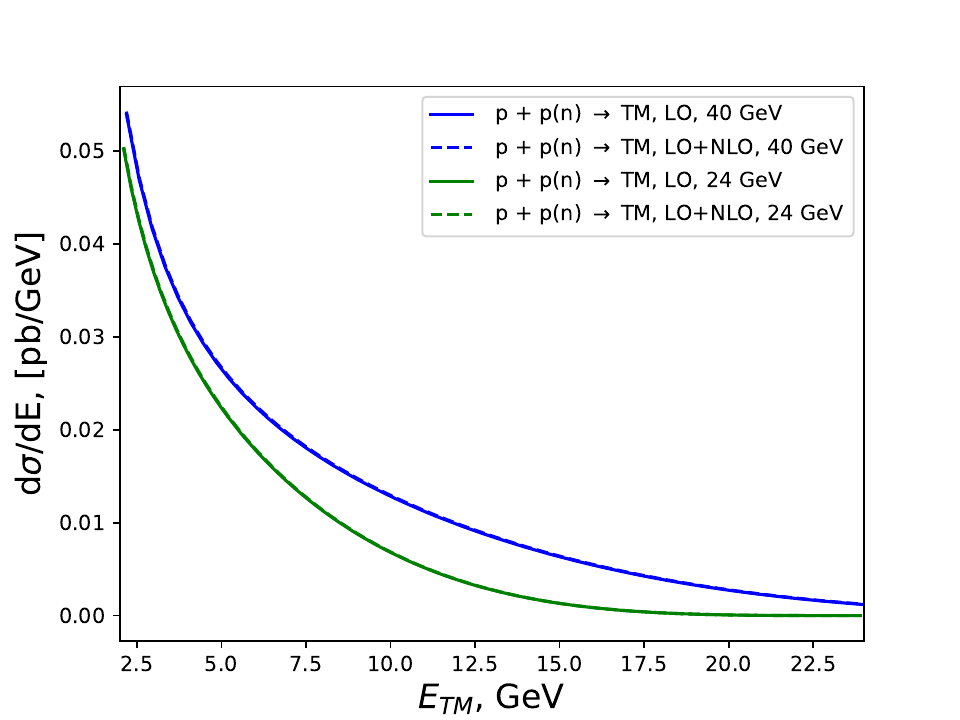}
  \caption{The differential cross section of the  $oT\!M$  production
          for DY processes induced by the $pp$ and $pn$ scattering at
          proton beam energies $E_{beam}=40$~GeV and $E_{beam}=24$~GeV. 
          Lower line corresponds to the  $oT\!M$  production for
          $E_{beam}=24$~GeV proton beam. Dashed lines from the  (LO+NLO) 
          calculations practically overlap those from the LO ones.}
	\label{cs_n_TM}
\end{figure}

The yield of $T\!M$s produced 
in the DY reaction with proton scattered off a nuclear
target is given by 
\eq 
N_{T\!M} \simeq n \frac{\rho_{T} N_A}{A} L_{T}
 \sigma_{DY}(p+(A,Z)\to T\!M + X), 
\label{N_TM}
\en
where $n$ is the number of POT, $A$ and $Z$ are the atomic number and
charge of the nucleus, $N_A$ is the Avogadro's number, $\mbox{POT}$
is the number of protons accumulated on target, $\rho_{T}$ is
the target density, $L_T$ is the thickness of the target.
Inclusive integral cross section of
$T\!M$ production in proton scattering 
off nuclear target $(A,Z)$ is given by 
\eq
& &\sigma_{DY}(p+(A,Z)\to T\!M + X)
\nonumber\\
&=& Z \sigma_{DY}(p+p\to T\!M + X)
\nonumber\\
&+&(A-Z)\sigma_{DY}(p+n\to T\!M + X)\,. 
\en
Based on the collinear QCD factorization picture
had\-ro\-nic DY cross section is given by the convolution
of partonic-level DY cross section and 
partonic distribution functions for
nuclear (PDFs)~\cite{Qiu:2003cg,Kovarik:2015cma}.
Perturbative $\alpha_s$ expansion of the partonic-level DY cross section 
including LO and NLO terms reads
$d\sigma_{DY}^{ab}=d\sigma_{0}^{ab}+(\alpha_s(Q^2)/\pi) d\sigma_{1}^{ab} +
{\cal O}(\alpha_s^2)$~\cite{Shimizu:2005fp},  
where LO contribution~\cite{dEnterria:2023yao} is given by 
\begin{align}
  d\sigma_{0}^{ab} &= \int_{z/x_1}^{1} dx_1\int_{z/x_2}^{1} dx_2
  \, \Big(f_{q/N}(x_1,\mu_F)f_{\bar{q}/N}(x_2,\mu_F)
\nn \\&+f_{\bar{q}/N}(x_1,\mu_F)f_{q/N}(x_2,\mu_F)\Big) 
\nn \\& \times
\delta\left(1-\frac{m^2_{T\!M}}{\hat{s}}\right)
\hat{\sigma}_{q\bar{q}\to T\!M}(\hat{s},\mu_F)
\,.
\end{align}
Here, 
$\hat{s}=x_1x_2 S$ and
$S=(p_1+p_2)^2 \simeq 2 m_N E_{beam}$ are 
the partonic and hadronic total energies, 
$f_{q(\bar{q})/N}(x_1,\mu_F)$ are quark (antiquark) PDFs, 
$x_i$ is the longitudinal fraction of hadron momentum,
$\mu_F$ is the factorization scale which equal to energy
of $T\!M$, $z=m_{T\!M}^2/S$, 
$\hat{\sigma}_{q\bar{q}\to T\!M}(\hat{s},\mu_F) =
\alpha\epsilon_{T\!M}^2  Q_q^2/m^2_{T\!M}$
is partonic-level cross section,
and $Q_q$ is the quark charge. 

The DY cross sections at LO and NLO 
are calculated numerically by using CalcHep~\cite{Pukhov:1999gg}
and CTEQ61L~\cite{Pumplin:2002vw}
parametrization for PDFs of protons and neutrons. The corresponding
partonic-level diagrams are presented in Figs.~\ref{qqTM}-\ref{Diag_qg}. 
In the case of NLO calculations  the upper cut of 1 GeV is used 
for the $T\!M$ transverse momentum. The size of the NLO correction
to the LO result is demonstrated in Fig.~\ref{cs_n_TM}
for the $T\!M$ energy $E_{T\!M}$,
Here we show the results for cross section for two
proton beam energies: 40 GeV and 24 GeV. Consideration of
24 GeV beam is motivated by operation of the
CERN PS accelerator at this energy. 
One can see that the NLO correction play a weak role for the $T\!M$ production
for $E_{T\!M} $ (see Fig.~\ref{cs_n_TM}). 
As the $pp$ and $pn$ differential cross sections have very similar
behavior, with a good accuracy
we can approximate the total DY cross section as
$\sigma_{DY}(p + (A,Z)\to T\!M + X) \simeq A \times
\sigma_{DY}(p+N\to T\!M + X)$, where $N$ is nucleon.
It helps to estimate the DY dimuon production for different
types of target nuclei in a simple way.

\par A promising  technique for discovering of $T\!M$ is based on
the observation of its decays in flight from the state 
$n^3S_1$ into the $e^+ e^-$ pair at some distance from the foil target.  
The cross section for the $T\!M$ dissociation in the target was evaluated 
in Refs.~\cite{Mrowczynski:1985qt,Holvik:1986ty}.
It was also found that for the $T\!M$ gamma-factor $\gamma>6$
the dissociation ($\sigma_{\rm dis}$)  and excitation ($\sigma_{\rm exc}$)
cross sections do not dependent on the atom momentum \cite{Nemenov:1981kz}. In addition,
the $\sigma_{\rm dis}$ has a simple power dependence on the target nuclei charge 
$\sigma_{\rm dis} = 1.3 \, Z^2 \, 10^{-23}$ cm$^{-2}$.
Besides, one needs to point that  for excited
states $n ^3S_1$ the $\sigma_{\rm dis} \sim n^3$, resulting in a suppression
of the  excited $T\!M$ atom yield.  

\begin{center}
	
	\begin{table}[t]
		\begin{center}
		  \caption{Target parameters used
                    for estimate a yield of $T\!M$ 
                    passing through a metal foil including the number
                    of $T\!M$ generated in foil at 24 GeV
                    ($N^{24}_{T\!M}$) and 40 GeV ($N^{40}_{T\!M}$).}
			\vspace*{.1cm}
			\def\arraystretch{1.25}
			\begin{tabular}{|c|c|c|c|c|}
				\hline
                        $A\,\,\, (Z) \,\, [{\rm Name}]$ & $\rho g $ cm$^{-3}$
                        & $L_{T\!M}^{\rm eff}$ mm
                        &$N^{24}_{T\!M}$ & $N^{40}_{T\!M}$\\
			\hline
			56 (26) [Fe] & 	7.874 &0.013& 4250& 5811\\
			48 (22) [Ti] & 	4.540 &0.0278& 2444&3350\\
			27 (13) [Al] & 	2.699 &0.075&1456 &1996\\
			9 \,\,(4) \,\,[Be] & 	1.848  &0.388&993 &1360\\
			\hline
			\end{tabular}
			\label{tab:1}
		\end{center}
	\end{table}
	
\end{center}

\begin{figure}[t]
 \vspace*{-.25cm}
  \includegraphics[width=0.48\textwidth,clip]{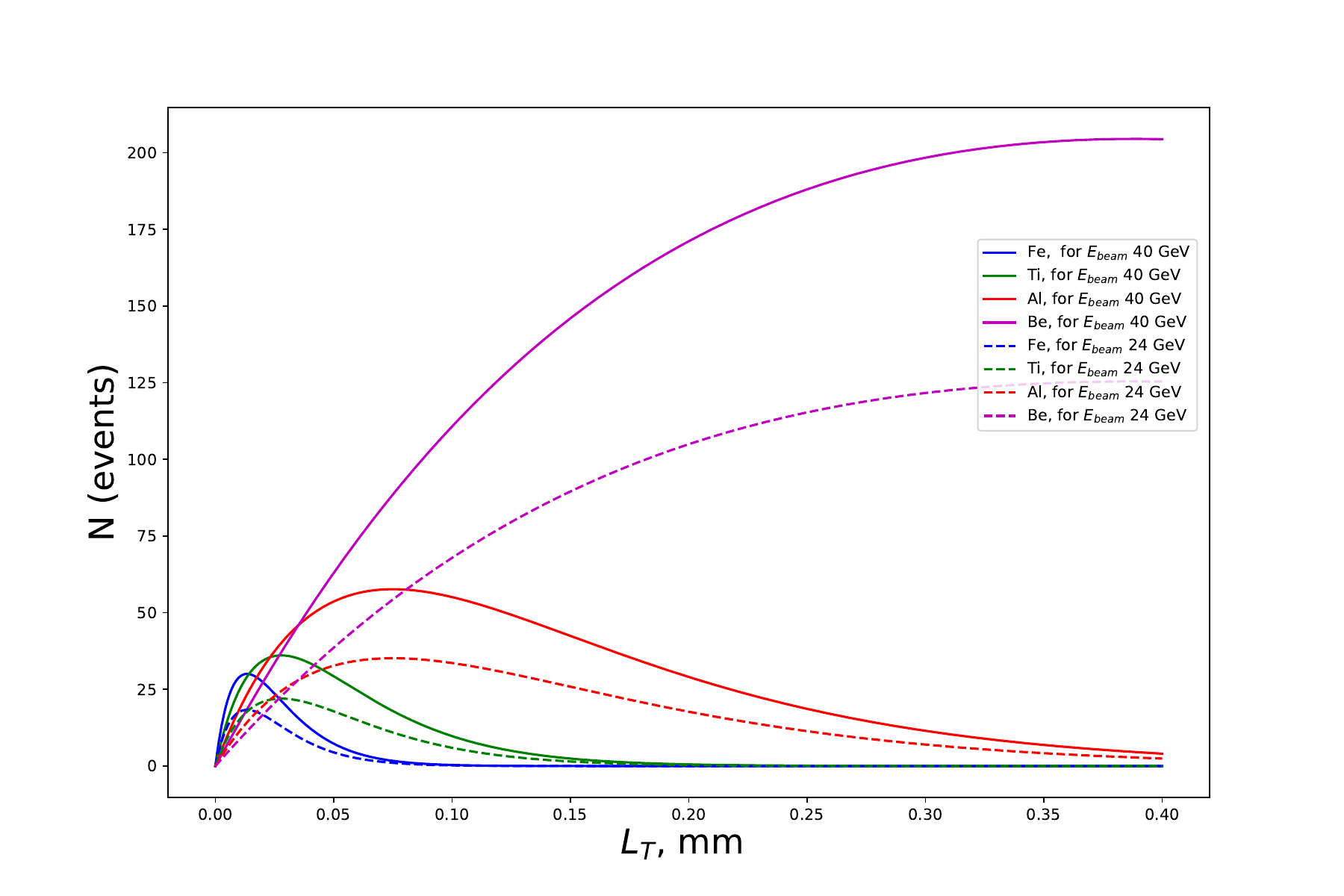}

         \vspace*{-.25cm}
         \caption{The yield of triplet $T\!M$ vs the 
          thickness of different target foils for $E_{T\!M} > 2$ GeV
          and $4 \times 10^{16}$ POT for 40 GeV (solid lines)
          and 24 GeV (dased lines) energies of proton beams.}
	\label{tagetfoilyield}

        \vspace*{-.1cm}
        
\includegraphics[width=0.48\textwidth,clip]{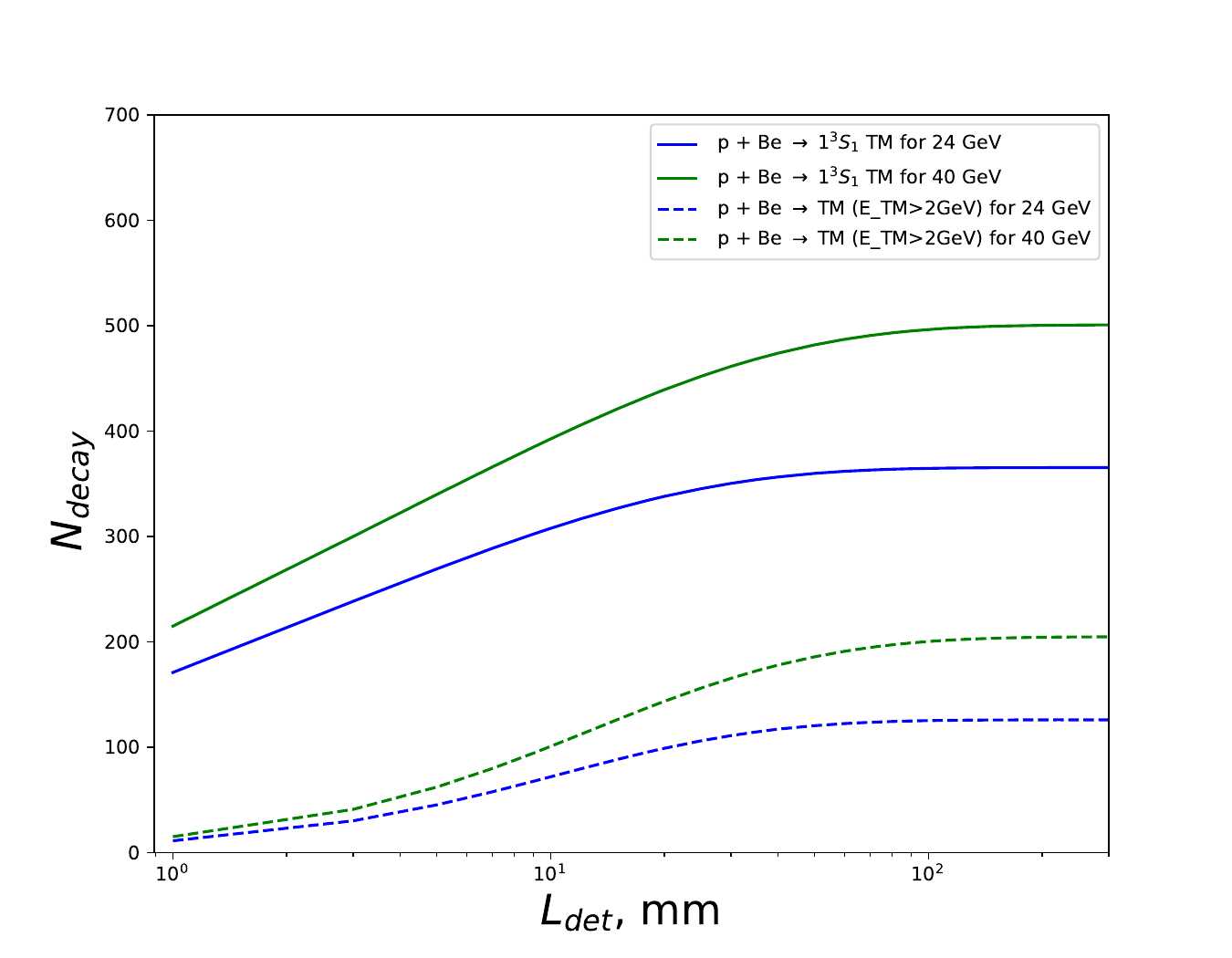}     
\caption{The  number $N_{decay}$ of  $oT\!M \to \ee$ decays 
at the distance $0 < L <L_{det}$ from the target 
as a function of   $L_{det}$. The calculations are done
for the $^{4}$Be foil of  0.388 mm thick
by  taking into account the $oT\!M$ absorption in the target,
for $4 \times 10^{16}$ POT,  beam energies of 24 and 40 GeV,
and the $oT\!M$ energies $E_{T\!M}> 0$ (solid curves)
and $E_{T\!M}> 2$ GeV
(dashed curves).}
\label{distance_decay_detection}	
\end{figure}

\vspace*{-.45cm}

The effective thickness $L_{T\!M}^{\rm eff}$ for $T\!M$ to pass
thought the target foil without  breaking up is given by 
$L_{T\!M}^{\rm eff} = A/(\rho_T N_A  \sigma_{dis})$. 
The probability that $T\!M$ suffers a collision between $l$
to $l+dl$ is given by 
$p(l) \, dl = \exp(-l/L_{T\!M}^{\rm eff})/L_{T\!M}^{\rm eff}
\, dl$~\cite{Lechner:2018bzx}.  
Therefore, the probability for passing through the target can be
approximated as  $W(L_{T\!M}) = \exp(-L_{T\!M}/L_{T\!M}^{\rm eff})$,  
where $L_{T\!M}$ is the experimental target thickness~\cite{Nemenov:1981kz}.
The dependence of the $T\!M$  yield (for $E_{T\!M}$>2 GeV)
on the foil thickness is presented in Fig.~\ref{tagetfoilyield}.  
The maximum $T\!M$ yield   corresponds to the foil thickness equal
to the effective  $T\!M$  free-pass length in the target.
The effective thickness $L_{T\!M}^{\rm eff}$ is inversely proportional
to the density and $Z^2$ of nuclei of the target. Thus, 
the low-density foil material is preferable to  get the highest yield. 
Indeed, as one can see in Fig.~\ref{tagetfoilyield},
for having the highest yield of $oT\!M$ atoms
with a reasonable detection energy $E_{T\!M} > 2$ GeV, 
the using of the beryllium target foil is more preferable. 
The  full number  of surviving  true muoniums
for different types of target foils with  their specific thickness value
$L^{\rm eff}_{T\!M}$ is presented in Table.~\ref{tab:1}.
As one can see, for the  $^4$Be foil with the thickness of
$L^{\rm eff}_{T\!M}\simeq 0.388$ mm, resulting in the suppression
of the $oT\!M$  yield by $\simeq 1/e$, we can obtain better yield of $T\!M$ atoms through foil.  One can get 
the  full yield of  $\sim 1000$ and $\sim 1400$ $1 ^3S_1$ states
for 24 GeV and 40 GeV beam energy, respectively, and $n=4\times 10^{16}$ POT. 
This results in the total production of $\simeq 360$  and
$\simeq 510$ $\otm$ atoms, respectively, that passed through the foil.
The states of $T\!M$ with the energy $E_{oT\!M} > 2$ GeV, we will have $\simeq 125$  and
$\simeq 187$ $\otm$ atoms for 24 GeV and 40 GeV beam energy, respectively. 
Using these yields of $oT\!M$ and its  lifetime $\tau=1.81$ ps, 
we present in Fig.~\ref{distance_decay_detection} 
the  number $N_{decay}$ of  $oT\!M \to \ee$ decays at the distance
$0 < L <L_{det}$  from the $^4$Be foil as a function of $L_{det}$ estimated
by taking into account the decay probability as
$P_{decay}=1-\exp(-L_{det}/(\tau \beta \gamma))$, 
where $\beta\gamma$ is the Lorentz factor corresponding to the momentum of $T\!M$ state. 

\par For  the same statistics, the yield of $T\!M$s
in the long-lived $2 ^3S_1$ state with $\tau=14.5$ ps is  estimated 
to be  $\sim 120$ and $\sim 170$ events for the 24 GeV and 40 GeV
beam energy, respectively. Besides, the $T\!M$ excited states can be
also produced  while its passing trough the foil, see recent
Ref.~\cite{Alizzi:2024wxp}, and also~\cite{Banburski:2012tk}.
The excitation of the $T\!M$ state is suppressed 
in force that characteristic (dimensionless) momentum for excitation
is $\tilde{q}\sim \alpha m_\mu a_B=2$ ($a_B$
is the radius of the first Bohr orbit in $T\!M$) wherein transferred
momentum corresponding to the Thomas-Fermi screening scale
$q_c=1.09 \times 10^{-2}Z^{1/3}$ is equal 0.017 for
$^{4}$Be target~\cite{Alizzi:2024wxp}.
It means that the $T\!M$ atom must pass very close to
the nucleus to be excited and all effects of interaction
with the Coulomb potential of the nucleus are gone. 
For mildly relativistic muonium, the  cross section
for  transitions into different atomic states can be  estimated by solving
the transport equation~\cite{Alizzi:2024wxp}.
In particular, the result for the cross section
for transition in $^{4}$Be foil is 
$\sigma(1^3S_1\to 2P)=5.322 \times 10^{-23}$ cm$^2$ ~\cite{Alizzi:2024wxp}.
Thus, one can produce $\sim 200$ excited atomic states.
However, for a foil with thickness equal to $L^{\rm eff}_{T\!M}$
the yield of excited states will be close to zero due to the
large deexcitation cross section for all excited states.  
To ensure the yield of excited states through a foil, one needs to decrease
thickness of the target and increase the number of POT.
Finally, we note that the prediction error for the $oT\!M$
yield is connection with value of cross section of $T\!M$ dissociation
in matter. Difference phenomenology formula  cross section of $T\!M$
dissociation in $^4$Be or other matter and quantity from
Ref.\cite{Alizzi:2024wxp} is order 5 $\div$ 15 percents. 

\par 
It is worth mentioning, that in the case of the proton
scattering off a foil one could also produce $\eta$ and $\eta^\prime$ mesons
in the exclusive reaction $p+Z \to \eta (\eta^{\prime})+X$.
For these mesons the branching ratios
of the $T\!M$ production have been estimated to be  
$Br(\eta \to T\!M +\gamma)< 4.7 \times 10^{-10}$ and $Br(\eta^\prime
\to T\!M +\gamma)< 3.7 \times 10^{-11}$, see  Ref.~\cite{Ji:2018dwx}. 
With the use of the Geant4 simulation~\cite{GEANT4:2002zbu}
for $4\times 10^{16}$ protons on the $^4$Be foil target 
and beam energies 24 GeV and 40 GeV we obtained  $\sim 2\times 10^4$ and
$\sim 10^3$
as an estimate for the number of $\eta$ and $\eta^\prime$ mesons decaying
into $T\!M$ state and photon, respectively. 
At the same time, decays involving muonium production will occur at a 
different distance compared to the decays of $1^3S_1$ states of $T\!M$
produced in the Drell-Yan process. These decays provide an additional
source for detecting $T\!M$ atomic states.

\par 
An experiment to search for the $T\!M$ production could be performed with
an apparatus specifically designed to search for the $oT\!M \to \ee$ decays
of $oT\!M$s  escaping from the target to vacuum, which is similar to 
the DIRAC setup used for measurements of the $\pi^+ \pi^-$ atom
lifetime~\cite{Nemenov:1984cq}. 
Briefly, it consists of a magnetic  spectrometer with two-arms for detection
of decay electrons and positrons. Each arm is equipped with a high-precision
tracker with low material content followed by an electromagnetic (e-m)
calorimeter in which e-m showers from electrons and photons are detected
and measured. Due to the relatively
large opening angle of decay pairs, $\Theta_{\ee} \simeq 2m_\mu/E_{T\!M}$,
$20 \lesssim \Theta_{\ee} \lesssim 100$ mrad, for the $T\!M$ energy range
$2 \lesssim E_{T\!M} \lesssim 10$ GeV (see Fig.~\ref{tagetfoilyield}) 
the displaced form 
the target vertex of the $oT\!M$ along $Z$-axis can be reconstructed with 
an accuracy $\delta Z \simeq$ a few mm. The discovery of $T\!M$ would be
the observation of an excess of the $\ee$ events with the displaced vertex
in vacuum and  the invariant mass and  decay time consistent 
with the mass and lifetime of the $oT\!M$ state compared to
the expected background. The main background source of fake
$oT\!M$ events is expected from kaons produced by primary protons in 
the target and followed by their $K^\pm,K_{S,L} \to \ee + X$ decays downstream
of the target. As preliminary analysis shows, the decays accompanied by
an additional charged track, e.g., such as in decay chain
$K^\pm \to \pi^0 e^\pm \nu; \pi^0 \to e^+ e^- \nu$ can be suppressed by
the requirement of no additional track in event pointed to the $\ee$ vertex, 
while such decays as $K_L\to \ee \gamma$ can be rejected kinematically.
In order to obtain the best sensitivity of the $T\!M$ search, a compromise
should be found between the $oT\!M$ yield, background level, accuracy of
the decay vertex reconstruction, and  the energy and intensity of the proton
beam. An experiment with the proton beam energy
around $\sim 20$ GeV, available, e.g., at CERN PS or JINR (Dubna),
seems is more preferable. 
A more complete description of the setup design and analysis of the
sensitivity will be published elsewhere~\cite{proposal}.

In conclusion, in this Letter we consider a new mechanism of the
$T\!M$ formation in a proton FT experiment  
due to the Coulomb interaction of muons from  $\mu^+ \mu^-$ pairs 
 produced in the DY scattering of
medium-energy protons off nuclei of a thin foil target. 
The experimental signature of the $T\!M$  formation is the observation
of displaced $\ee$ pairs downstream of the target from the decays
of triplet $T\!M$s escaping from the foil. 
We present estimates of the $T\!M$ yield, and show that it
is at a measurable level. 
A  brief discussion of a possible experiment, which could
lead to the $T\!M$ discovery in a near future is presented. 

\begin{acknowledgments}

We thank S. Brodsky, P. Crivelli, and N. Krasnikov for interest 
to our work and useful discussions. We thank A. Ivanov and M. Kirsanov
for simulations with  GEANT4.
This work was funded by ANID$-$Millen\-nium Program$-$ICN2019\_044 (Chile), 
by FONDECYT (Chile) under Grant No. 1240066. 
The work of A. S. Zh. is supported by the Foundation for
the Advancement of Theoretical Physics and Mathematics "BASIS".
  
\end{acknowledgments}

\end{document}